\documentstyle[12pt,psfig]{article}

\begin{document}
\pagenumbering{arabic}
\begin{center}
{\Large \bf Study of the neutron star structure in strong magnetic fields
   including the anomalous magnetic moments}\\
\vspace{0.8cm} {Guangjun~Mao$^{1,2)}$, Akira~Iwamoto$^{1)}$ and
Zhuxia~Li$^{3)}$}
\end{center}
\begin{flushleft}
\hspace{2.3cm} {\it $^{1)}$Japan Atomic Energy Research Institute \\
\hspace{2.5cm}    Tokai, Naka, Ibaraki 319-1195, Japan}\\
\hspace{2.3cm} {\it $^{2)}$Institute of High Energy Physics,
                           Chinese Academy of Science \\
\hspace{2.5cm}      P.O. Box 918(4), Beijing 100039, P.R.
China\footnote{Permanent address \\ \indent \hspace{0.2cm}
 e-mail: maogj@mail.ihep.ac.cn}}\\
\hspace{2.3cm} {\it $^{3)}$China Institute of Atomic Energy \\
\hspace{2.5cm}  P.O. Box 275(18), Beijing 102413, P.R. China}
\end{flushleft}
%\date{\today}
\begin{abstract}
\begin{sloppypar}
We study the effects of strong magnetic fields on the neutron star
structure. If the interior field of a star is on the same order of
the surface field currently observed, the influences of the magnetic
field on the star mass and radius are negligible. If one assumes that
the internal magnetic field can be as large as that estimated from the scalar
virial theorem, considerable effects can be induced. The maximum mass of
stars is arisen substantially while the central density is largely suppressed.
For two equal-mass stars the radius of the magnetic star can be larger by
about 10\% $\sim$ 20\% than the nonmagnetic star.

\end{sloppypar}
\end{abstract}
\newcounter{cms}
\setlength{\unitlength}{1mm}
\newpage
  \begin{center}
 {\bf I. INTRODUCTION}
  \end{center}
 \begin{sloppypar}
 It is well known that the structure of neutron stars is mainly determined by
 the nuclear equation of state (EOS) built on the strong interactions. Recent
 observations have indicated that large magnetic fields are presented at the
 surface of neutron stars (Michel 1991; Rothschild, Kulkarni \& Lingenfelter
  1994; Kouveliotou et al. 1998; Woods et al. 1999).
 The dipole fields
 inferred from the spin-down rates of neutron star rotations can be up to
 $10^{15}$ G (for a recent review, see Reisenegger (2001)). The strength of
 magnetic fields in the interiority of stars remains unknown. According to the
 scalar virial theorem (Shapiro \& Teukolsky 1983; Lai \& Shapiro 1991)
 the maximum interior field strength could reach $\sim 10^{18}$ G for a star
 with $R \approx  10$ km and $M \approx 1.4$ $M_{\odot}$. Even larger fields
 may be expected in the core of neutron stars. Such high fields will
 certainly play a role when one evaluates the EOS of neutron-star matter.
 Consequently, it may causes considerable effects on the structure of
 neutron stars.

 The same problem has been addressed on the magnetic white dwarfs where a
 surface magnetic field on the order of $10^{5}$ $\sim$ $10^{9}$ G and an
 interior field of $10^{9}$ $\sim$ $10^{13}$ G are estimated. Both the
 earlier work of Ostriker \& Hartwick (1968) and recent calculations of
 Suh \& Mathews (2000) predicted an increase of white dwarf radii in the
 presence of internal magnetic fields. In the case of white-dwarf binary
 system of LB11146 (PG 0945+245), if one assumes an equal mass of $M=0.9$
 $M_{\odot}$ for the magnetic one and nonmagnetic one, an interior field
 of $B \approx 0.5 B_{c}^{e}$ (here $B_{c}^{e}=4.414 \times 10^{13}$ G is the
 electron critical field) would cause the radius of the magnetic white dwarf
 be larger by about 10\% than that of the normal star (Suh \& Mathews 2000).

 Theoretical investigation of ideal noninteracting neutron-proton-electron
 ($n$-$p$-$e$) gas and interacting pure neutron matter under large magnetic
 fields has been carried out by Suh \& Mathews (2001) and Vshivtsev \& Serebryakova
  (1994). Recently, Br\"{u}ckner-Hartree-Fock calculations of spin polarized
  asymmetric nuclear matter have been performed (Vidana \& Bombaci 2002).
 Based on the meson field theory several authors have incorporated strong
 magnetic fields into the equation of state of a dense $n$-$p$-$e$ system
 under the beta equilibrium and the charge neutrality conditions
 (Chakrabarty, Bandyopadhyay \& Pal 1997, Paper I;
  Broderick, Prakash \& Lattimer 2000, Paper II).
  Evident changes on the EOS have been found. It was demonstrated that the nucleon
 anomalous magnetic moments (AMM) play a significant role (Paper II)
 which may overwhelm the softening of the EOS caused by Landau
 quantization (Paper I) to the stiffening.
 In the mean time, a dramatic increase of the proton fraction with the
 increase of the magnetic field was exhibited. However, our previous
 calculations (Mao et al. 2002) showed that with the AMM of
 nucleons and electrons taken into account the proton fraction was found
 to never exceed the field free case. Extremely strong fields would lead to
 a pure neutron matter rather than a proton-rich matter. In this work we will
 study the effects of large magnetic fields on the neutron star structure.
 We will examine the EOS of a dense $n$-$p$-$e$ system with the AMM
 of both nucleons and electrons taken into account.
 One may argue that the electron self-energy may not change substantially
 in magnetic fields when high-order terms are taken into account. However,
  a systematic incorporation of high-order contributions beyond the AMM term
  is not yet clear, which will be the topic of our forthcoming works. Here the
  effects of magnetic fields on different particles within the considered system
  are treated on an equal footing. The developed EOS will
 be subsequently applied to investigate the structure of neutron stars with
 strong internal fields. The paper is organized as follows: a relativistic
 mean-field theory approach for  dense neutron star matter is described
 in Section 2. In Section 3 we present the numerical results.
 A brief summary and outlook will be finally given in Section 4.

  \begin{center}
 {\bf II. THEORETICAL FRAMEWORK}
  \end{center}

We consider a neutron-star matter consisting of neutrons, protons and electrons
interacting through the exchange of $\sigma$, $\omega$ and $\rho$ mesons in
the presence of a uniform magnetic field $B$ along the $z$ axis. The Lagrangian
density can be written as (Serot \& Walecka 1986)
 \begin{eqnarray}
{\cal L} &=& \bar{\psi} \lbrack i\gamma_{\mu}\partial^{\mu}
   - e \frac{1+\tau_{0}}{2} \gamma_{\mu}A^{\mu} -\frac{1}{4}\kappa_{b}\mu_{N}
   \sigma_{\mu\nu}F^{\mu\nu}
  - M_{N} + {\rm g}_{\sigma}\sigma
   - {\rm g}_{\omega}\gamma_{\mu}\omega^{\mu}
    -\frac{1}{2}{\rm g}_{\rho}\gamma_{\mu} \mbox{\boldmath $\tau$}\cdot
    {\bf R}^{\mu} \rbrack \psi \nonumber \\
 && + \bar{\psi}_{e} \lbrack i \gamma_{\mu}\partial^{\mu} - e \gamma_{\mu}
    A^{\mu} - \frac{1}{4}\kappa_{e}\mu_{B}\sigma_{\mu\nu}F^{\mu\nu}
    -m_{e} \rbrack \psi_{e}
  + \frac{1}{2}\partial_{\mu}\sigma \partial^{\mu}\sigma - U(\sigma)
  \nonumber \\
 &&   -\frac{1}{4}\omega_{\mu\nu}\omega^{\mu\nu} + \frac{1}{2}m_{\omega}^{2}
    \omega_{\mu}\omega^{\mu}
  - \frac{1}{4}{\bf R}_{\mu\nu}\cdot{\bf R}^{\mu\nu} + \frac{1}{2}
 m_{\rho}^{2}{\bf R}_{\mu}\cdot {\bf R}^{\mu},
 \end{eqnarray}
 and $U(\sigma)$ is
 the self-interaction part of the scalar field (Boguta \& Bodmer 1977)
 \begin{equation}
 U(\sigma)=\frac{1}{2}m_{\sigma}^{2}\sigma^{2}+ \frac{1}{3}b({\rm g}_{\sigma}
 \sigma )^{3} + \frac{1}{4}c({\rm g}_{\sigma}\sigma )^{4}.
 \end{equation}
 In the above expressions $\psi$ and $\psi_{e}$ are the Dirac spinors of the
 nucleon and electron; $\sigma$, $\omega_{\mu}$, ${\bf R}_{\mu}$ represent
 the scalar meson, vector meson and vector-isovector meson field, respectively.
 $A^{\mu} \equiv (0,0,Bx,0)$ refers to a constant external magnetic field.
 Here the field tensors for the omega, rho and magnetic field are given in terms
 of their potentials by
   \begin{eqnarray}
 && \omega_{\mu\nu}=\partial_{\mu}\omega_{\nu} - \partial_{\nu}\omega_{\mu}, \\
 && {\bf R}_{\mu\nu}=\partial_{\mu}{\bf R}_{\nu} - \partial_{\nu}{\bf R}_{\mu}, \\
 && F_{\mu\nu}=\partial_{\mu}A_{\nu} - \partial_{\nu}A_{\mu},
   \end{eqnarray}
 and $\sigma_{\mu\nu}=\frac{i}{2} \lbrack \gamma_{\mu}, \gamma_{\nu} \rbrack$.
 $\mbox{\boldmath $\tau$}$ is the isospin operator of the nucleon and $\tau_{0}$
 is its third component. $M_{N}$ and $m_{e}$ are the free nucleon mass and
 electron mass and $m_{\sigma}$, $m_{\omega}$, $m_{\rho}$ are the masses of the
 $\sigma$-, $\omega$- and $\rho$-meson.
 $\mu_{N}$ and $\mu_{B}$ are the nuclear magneton of nucleons and Bohr
 magneton of electrons; $\kappa_{p}=3.5856$, $\kappa_{n}=-3.8263$ and
 $\kappa_{e}=\alpha / \pi$ are the coefficients of the AMM for protons,
 neutrons and electrons (Greiner \& Reinhardt 1994), respectively.
 The third set of parameters presented by Glendenning \& Moszkowski (1991)
  is used as the nucleon coupling strengths.
 It gives ${\rm g}_{\sigma}=8.7818$, ${\rm g}_{\omega}=8.7116$,
 ${\rm g}_{\rho}=8.4635$, $b{\rm g}_{\sigma}^{3}=27.9060$, $c{\rm g}_{\sigma}
 ^{4}=-14.3989$. This yields a binding energy $B/A=-16.3$ MeV, saturation
 density $\rho_{0}=0.153$ fm$^{-3}$ and bulk symmetry energy $a_{sym}=32.5$
 MeV.

 Let us first consider nucleons. The Dirac equation for the nucleons in a
 uniform magnetic field can be written as
  \begin{equation}
 \left[ i\gamma_{\mu}\partial^{\mu}
   - e \frac{1+\tau_{0}}{2} \gamma_{\mu}A^{\mu} -\frac{1}{4}\kappa_{b}\mu_{N}
   \sigma_{\mu\nu}F^{\mu\nu}
  - M_{N} + {\rm g}_{\sigma}\sigma
   - {\rm g}_{\omega}\gamma_{\mu}\omega^{\mu}
    -\frac{1}{2}{\rm g}_{\rho}\gamma_{\mu} \tau_{0}
     R^{\mu}_{0} \right] \psi =0  .
  \end{equation}
  The general solutions of the above equation in the free space, i.e.,
 in the absence of the meson fields, are discussed in Appendix A. The
 corresponding solutions in the neutron-star matter can be directly written out by
 replacing the free quantities with the effective ones.  The positive energy of
 the protons in the Fermi sea $\left(E_{\nu,S}^{p} \right)_{+}$ and the negative
 energy of the protons in the Dirac sea $\left( E_{\nu,S}^{p} \right)_{-}$
 read as
   \begin{eqnarray}
 && \left( E_{\nu,S}^{p} \right)_{+} = \left\{ \left[ \left( \sqrt{m^{*2}
   + 2eB\nu} + S\Delta \right)^{2} + p_{z}^{2} \right]^{1/2}
   + {\rm g}_{\omega}\omega_{0} + \frac{1}{2}{\rm g}_{\rho}R_{0,0} \right\}, \\
 && \left( E_{\nu,S}^{p} \right)_{-} = - \left\{ \left[ \left( \sqrt{m^{*2}
   + 2eB\nu} + S\Delta \right)^{2} + p_{z}^{2} \right]^{1/2}
   - {\rm g}_{\omega}\omega_{0} + \frac{1}{2}{\rm g}_{\rho}R_{0,0} \right\}.
   \end{eqnarray}
   The positive-energy and negative-energy spectra of the neutrons are
   \begin{eqnarray}
 && \left( E_{S}^{n} \right)_{+} = \left\{ \left[ \left( \sqrt{p_{x}^{2}
   + p_{y}^{2} + m^{*2}} + S\Delta \right)^{2} + p_{z}^{2} \right]^{1/2}
   + {\rm g}_{\omega}\omega_{0} - \frac{1}{2}{\rm g}_{\rho}R_{0,0} \right\}, \\
 && \left( E_{S}^{n} \right)_{-} = - \left\{ \left[ \left( \sqrt{p_{x}^{2}
   + p_{y}^{2} + m^{*2}} + S\Delta \right)^{2} + p_{z}^{2} \right]^{1/2}
   - {\rm g}_{\omega}\omega_{0} - \frac{1}{2}{\rm g}_{\rho}R_{0,0} \right\}.
   \end{eqnarray}
 Here $\Delta = -\frac{1}{2}\kappa_{b}\mu_{N}B$; $ S=\pm 1$ for spin-up and
  spin-down particles. $\nu$ is the quantum number of Landau levels for
  charged particles (Landau \& Lifshitz 1977).
  The positive energy of the anti-particle is just the negative of the negative
  energy of the particle, i.e., $\left(\bar{E}_{\nu,S}^{p} \right)_{+}=
  -\left(E_{\nu,S}^{p} \right)_{-}$, $\left( \bar{E}_{S}^{n} \right)_{+}= -
  \left( E_{S}^{n} \right) _{-}$ (Mao, St\"{o}cker \& Greiner 1999).

Since neutron stars are cold dense matter,
we shall perform numerical calculations at zero temperature. The general
chemical equilibrium is realized for valence particles.
 The chemical potentials of protons and neutrons are defined as
  \begin{eqnarray}
&&  \mu_{p} = \epsilon_{f}^{p} + {\rm g}_{\omega}\omega_{0} + \frac{1}{2}
    {\rm g}_{\rho}R_{0,0}, \\
 && \mu_{n} = \epsilon_{f}^{n} + {\rm g}_{\omega}\omega_{0} - \frac{1}{2}
    {\rm g}_{\rho}R_{0,0}.
  \end{eqnarray}
  They are related to the respective Fermi momenta via following equations:
  \begin{eqnarray}
 && \left( k_{f,\nu,S}^{p} \right)^{2} = \left( \epsilon _{f}^{p} \right)^{2}
    - \left( \sqrt{m^{*2} + 2eB\nu } + S \Delta \right)^{2},
    \label{pferm} \\
 && \left( k_{f,S}^{n} \right) ^{2} = \left( \epsilon_{f}^{n} \right) ^{2}
   - (m^{*} + S\Delta )^{2}.
  \end{eqnarray}
 In the above equations, the effective nucleon mass $m^{*}=M_{N}
  -{\rm g}_{\sigma}\sigma$. $\sigma$, $\omega_{0}$ and $R_{0,0}$ are the mean
  values of the scalar field, the time-like component of the vector field and
  the time-like isospin 3-component of the vector-isovector field in
  neutron-star matter, respectively. They are obtained by solving the non-linear
  equations of the meson fields
   \begin{eqnarray}
 && m_{\sigma}^{2}\sigma + b{\rm g}_{\sigma}^{3}\sigma^{2} + c{\rm g}_{\sigma}^{4}
    \sigma^{3} = {\rm g}_{\sigma}\rho_{S}, \\
 && m_{\omega}^{2}\omega_{0} = {\rm g}_{\omega}\rho, \\
 && m_{\rho}^{2}R_{0,0} = \frac{1}{2}{\rm g}_{\rho}\rho_{0,0}.
   \end{eqnarray}
 Here $\rho_{S}$, $\rho$ and $\rho_{0,0}$ are the scalar density, the time-like
 component of the vector density and the time-like isospin 3-component of the
 vector-isovector density contributed from the valence nucleons, that is, from the
 Fermi sea. In principle, there exist additional contributions
 stemming from the Dirac
 sea. Here they are neglected according to the {\em no-sea} approximation since
 the renormalization of the system under the external magnetic field is a problem
 to be solved. Thus, the contributed densities are
 $\rho_{S}=\rho_{S}^{p} + \rho_{S}^{n}$, $\rho = \rho_{0}^{p} + \rho_{0}^{n}$
 and $\rho_{0,0} = \rho_{0}^{p} - \rho_{0}^{n}$, with
 \begin{eqnarray}
&&  \rho_{S}^{p}= \frac{eBm^{*}}{2\pi^{2}}\sum_{S}\sum_{\nu}
    \frac{\sqrt{m^{*2}+2eB\nu}+S\Delta}{\sqrt{m^{*2}+2eB\nu}} \ln
    \left| \frac{k_{f,\nu,S}^{p} + \epsilon_{f}^{p}}{\sqrt{m^{*2}+2eB\nu}
    + S\Delta} \right|, \\
&&  \rho_{S}^{n}=\frac{m^{*}}{4\pi^{2}}\sum_{S}\left[ \epsilon_{f}^{n}
    k_{f,S}^{n} - (m^{*}+S\Delta)^{2} \ln \left| \frac{k_{f,S}^{n}+ \epsilon
    _{f}^{n}}{m^{*}+S\Delta} \right| \right] , \\
&&  \rho_{0}^{p}=\frac{eB}{2\pi^{2}} \sum_{S}\sum_{\nu}k_{f,\nu,S}^{p},
    \label{pden} \\
&&  \rho_{0}^{n} = \frac{1}{2\pi^{2}}\sum_{S}\left[ \frac{1}{3} \left(
    k_{f,S}^{n} \right) ^{3} + \frac{S\Delta}{2} \left( (m^{*}+S\Delta )
    k_{f,S}^{n} + \left( \epsilon_{f}^{n} \right) ^{2} (\arcsin \frac{m^{*}
    + S\Delta}{\epsilon_{f}^{n}} - \frac{\pi}{2}) \right) \right].
 \end{eqnarray}
 The summation of $\nu$ runs up to the largest integer for which $\left( k_{f,\nu,S}
 ^{p} \right)^{2}$ is positive. For spin-up protons $\nu$ starts from 1
 while for spin-down protons 0. It should be pointed out that here the so-called
 spin up and spin down are just relative notes since the wave functions are
 no more eigenfunctions of 3-component
  spin operator (see Appendix A), mainly attributed
 to the coupling of the spin to the magnetic field. The contributions of the
 protons and neutrons to the energy density read as
  \begin{eqnarray}
 \varepsilon_{p} &=& \frac{eB}{4\pi^{2}}\sum_{S}\sum_{\nu} \left[ k_{f,\nu,S}^{p}
    \epsilon_{f}^{p} + \left( \sqrt{m^{*2} + 2eB\nu} + S\Delta \right) ^{2}
    \ln \left| \frac{k_{f,\nu,S}^{p} + \epsilon_{f}^{p}}{\sqrt{m^{*2} + 2eB\nu}
    + S\Delta} \right| \right], \\
 \varepsilon_{n} &=& \frac{1}{4\pi^{2}}\sum_{S} \left\{
    \frac{1}{2}\left( \epsilon_{f}^{n} \right)^{3} k_{f,S}^{n} + \frac{2}{3}
    S\Delta \left( \epsilon_{f}^{n} \right)^{3} \left[ \arcsin \left(
    \frac{m^{*} + S\Delta}{\epsilon_{f}^{n}} \right) - \frac{\pi}{2} \right]
    \right.  \nonumber \\
  && \left.
    + \left[ \frac{1}{3}S\Delta - \frac{1}{4} \left( m^{*} + S\Delta\right) \right]
       \left[ (m^{*} + S\Delta) k_{f,S}^{n} \epsilon_{f}^{n}
     + (m^{*} + S\Delta)^{3} \ln \left| \frac{k_{f,S}^{n} + \epsilon_{f}^{n} }
       {m^{*} + S\Delta} \right| \right] \right\}.
  \end{eqnarray}

 In the $\beta$-equilibrium system, the electron is assumed to move freely in the
 strong magnetic fields. The wave functions of electrons are the same as that
 of protons in the free space except that the corresponding quantities are
 replaced by the electron ones. The energy spectrum of the electrons
 can be expressed as
   \begin{equation}
 \left( E_{\nu,S}^{e} \right)_{+} = \left[ \left( \sqrt{m_{e}^{2} + 2eB\nu}
  + S\Delta \right)^{2} + p_{z}^{2} \right] ^{1/2},
   \end{equation}
here $\Delta = - \frac{1}{2}\kappa_{e}\mu_{B}B$ and $\left( E_{\nu,S}^{e} \right)
 _{-} = - \left( E_{\nu,S}^{e} \right)_{+}$. The chemical potential of electrons
 $\mu_{e}=\epsilon_{f}^{e}$. Its relation to the electron Fermi momentum is
   \begin{equation}
  \left( k_{f,\nu,S}^{e} \right)^{2} = \left( \epsilon _{f}^{e} \right)^{2}
    - \left( \sqrt{m^{2}_{e} + 2eB\nu } + S \Delta \right)^{2}.
    \end{equation}
 The electron density is defined as
  \begin{equation}
  \rho_{0}^{e}=\frac{eB}{2\pi^{2}} \sum_{S}\sum_{\nu}k_{f,\nu,S}^{e}.
  \end{equation}
  The contribution of the electrons to the energy density reads as
  \begin{eqnarray}
 \varepsilon_{e} &=& \frac{eB}{4\pi^{2}}\sum_{S}\sum_{\nu} \left[ k_{f,\nu,S}^{e}
    \epsilon_{f}^{e} + \left( \sqrt{m^{2}_{e} + 2eB\nu} + S\Delta \right) ^{2}
    \ln \left| \frac{k_{f,\nu,S}^{e} + \epsilon_{f}^{e}}{\sqrt{m^{2}_{e} + 2eB\nu}
    + S\Delta} \right| \right].
  \end{eqnarray}

Finally, we obtain the energy density contributed from the neutron-star matter
  \begin{eqnarray}
 \varepsilon_{m} &=& \frac{1}{2}m_{\sigma}^{2}\sigma^{2} + \frac{1}{3}b \left(
    {\rm g}_{\sigma}\sigma \right)^{3} + \frac{1}{4}c \left( {\rm g}_{\sigma}\sigma
    \right) ^{4} \nonumber \\
    && + \frac{1}{2}m_{\omega}^{2}\omega_{0}^{2} + \frac{1}{2}m_{\rho}^{2}
       R_{0,0}^{2} + \varepsilon_{p} + \varepsilon_{n} + \varepsilon_{e}.
  \end{eqnarray}
 The total energy density of the system is given by
  \begin{equation}
 \varepsilon = \varepsilon_{m} + \frac{B^{2}}{8\pi},  \label{energy}
  \end{equation}
  where the last term is the contribution from
  the external electromagnetic field
  (Landau, Lifshitz \& Pitaevski$\breve{{\rm i}}$ 1984).
  In the charge neutral beta-equilibrated
  matter, the pressure of the system can be expressed as (Paper II)
  \begin{equation}
 p = \mu_{n} \rho - \varepsilon_{m} + \frac{B^{2}}{8\pi}. \label{pressure}
  \end{equation}
 The inclusion of the $B^{2}/8 \pi$ term in Eqs. (\ref{energy}) and
 (\ref{pressure}) is equivalent to introducing the electromagnetic-field
  tensor in the source term of the gravitational equation of general
  relativity (Bocquet et al. 1995; Bonazzola \& Gourgoulhon 1996; Cardall,
  Prakash \& Lattimer 2001). Numerical calculations are performed under
 the constraints of the charge neutrality $\rho_{0}^{p}=\rho_{0}^{e}$ and
 the $\beta$-equilibrium $\mu_{n}=\mu_{p}+\mu_{e}$ (Glendenning 1997).
 These two constraint
 equations together with three meson equations are solved self-consistently
 in an iteration procedure.

\end{sloppypar}
 \begin{center}
 {\bf III. NUMERICAL RESULTS}
 \end{center}
 \begin{sloppypar}
 Since there is no information directly available for the interior magnetic
 field of a star, we assume that the field varies from the surface to the
 center and adopt the following parametrization (Bandyopadhyay, Chakrabarty \&
 Pal 1997)
 \begin{equation}
 B(\rho / \rho_{0})=B_{surf} + B_{cent}\left[ 1 - \exp \left( - \beta
  (\rho / \rho_{0})^{\gamma}\right) \right],
 \end{equation}
 where the parameters are chosen to be $\beta=0.01$ and $\gamma=3$. We further
 set $B_{surf}=\alpha B_{c}^{e}$, $B_{cent}=\alpha \times 10^{4} B_{c}
 ^{e}$ and take $\alpha$ (which should not be confused with the
 fine structure constant.) as a free parameter to check the effects of
 different fields. In this configuration the magnetic field decreases from the
 center to the surface of a star. We further take the surface magnetic
 field in the range of $10^{13} \sim 10^{15}$ G which is in agreement with
 the values inferred from observations. Therefore, the applied magnetic
 field does not influence the stationary configuration of a spherical neutron
 star considered here (note that the stability of an axially symmetric neutron
 star in large average magnetic fields has been discussed recently (Broderick,
 Prakash \& Lattimer 2002; A.P.~Martinez, H.P.~Rojas \& H.J.M.~Cuesta 2003)).
 The EOS of neutron-star matter under strong magnetic
 fields is depicted in Fig.~1. It can be found that the equation of state
 becomes stiffer than the field free case at the main region of density.
 Stronger field leads to a stiffer EOS. For the very low density part the
 situation is just inverse. A softer EOS is obtained when the magnetic field
 is incorporated which can be more clearly seen from the upper panel of the
 figure.

 In Fig.~2 we separate the cases
 with and without the inclusion of the AMM effects. Here we have dropped the
 contributions from the electromagnetic field to the pressure and energy
 density and presented the EOS for the matter part. A uniform magnetic field
 from the surface to the center of neutron stars is assumed. Dashed line denotes
 the case of $B=50 B_{c}^{e}$ and neglecting the AMM effects. One can see that
 the EOS turns out to be softer than the field free case at small energy
 densities and approaches to it in the main domain. The inclusion of the AMM
 effects here practically leads to undistinguishable results since at such low
 field the anomalous magnetic moments do not play any role. The whole
 situation changes substantially when a strong magnetic field is considered,
 as depicted in the figure by the dash-dotted line and dash-dot-dotted line
 for $B=5 \times 10^{5} B_{c}^{e}$. In accordance with the findings in
 Paper I \& Paper II the EOS becomes softer compared to the case of $B=0$ when
 the AMM is neglected and stiffer when it is included. Therefore, the
 softening of the EOS at small energy densities exhibited in Fig.~1 is mainly
 due to the fact that at the surface region we have taken a relatively small
 magnetic field where the effects of the AMM are negligible. Under the
 employed parametrization the field increases with the increase of the
 density. The AMM terms gradually get into the game. This leads to a
 stiffening of the EOS as shown in the region of large energy densities.

 The structure of neutron stars can be obtained by applying the developed
 EOS to solve the Tolman-Oppenheimer-Volkoff (TOV) equation for a
 relativistic, spherical and static star (Shapiro \& Teukolsky 1983;
 Glendenning 1997).
 We obtain stable solutions for the TOV equation based on the EOS with
 the strong magnetic fields applied.
  Fig.~3 displays the gravitational masses of neutron stars
 as a function of central densities, i.e., the sequences of stars obtained
 under different circumstances of interior magnetic fields as indicated by
 different curves in the figure. Rather evident effects are induced by strong
 magnetic fields. The maximum mass of stars is increased
 drastically in the presence of strong fields.
 For the case of $\alpha=50$ the $M_{max}$ is
 enhanced by about 40\% compared to the field free case,
 which can nearly balance
 the effects of decreasing the maximum mass if the hyperon degrees of
 freedom are taken into account (Glendenning \& Moszkowski 1991). In the
 mean time the central density is largely suppressed with the increase of
 fields. This can be seen  from Fig.~1 since the equation of state
 becomes stiff when a strong magnetic field is buried in a star.
 Note that an enhancement of the maximum mass around 13\% -- 30\%
 has been reported (Bocquet et al. 1995; Bonazzola \& Gourgoulhon 1996;
 Cardall, Prakash \& Lattimer 2001) where the authors
 accomplished calculations for the structure of axisymmetric relativistic
 stars. The effects of strong magnetic fields were included in the
 stress-energy tensor of the gravitational equation while neglected in the input
 equations of state.

 We show in Fig.~4 the radii of neutron stars as a function of interior
 magnetic field strength. Three cases of fixed star masses with $M/M_{\odot}
 =1.4$, 1.6, and 1.8 are investigated. If the field is weak, as in the
 normal case, the heavier stars have smaller radii due to the effects of
 the gravitational force. However, the radius increases with the increase of
 the magnetic field strength to a large extent. When strong enough fields
 are presented, the heavier stars can even have larger radii since neutron
 stars with different internal magnetic fields may belong to different
 star sequences. Generally speaking, the radius of a magnetic star can be
 enhanced by about 10\% $\sim$ 20\% depending on the star mass compared to
 the nonmagnetic star of equal mass.

 We have also investigated the neutron star structure under a
 nonvaried field strength from the
 surface to the center of neutron stars. For $B=50B_{c}^{e}$ the relationship
 between the star mass and the central density is undistinguishable from that
 of the field free case. The enhancement of the star radii is found to be
 less than 2\%. It seems to be difficult to observe any effects of the magnetic
 field by measuring the star mass and radius if the magnetitude of  the
 interior field is on the same order of the surface field, though the
 influence of the surface field itself on the star properties should be
 pursued more closely in a model for matter below neutron drip
 (Baym, Pethick \& Sutherland 1971; Lai \& Shapiro 1991). Finally,
 in Fig. 5 we depict the mass-radius relations of different
 neutron-star sequences with the magnetic fields considered.
 In order to describe the surface region a model for non-uniform
 matter at low densities (Shen 2002) should be utilized, which
 will be taken into account in our future investigations.

 \end{sloppypar}
 \begin{center}
 {\bf IV. SUMMARY AND OUTLOOK}
 \end{center}
 \begin{sloppypar}
 Within a relativistic field theory approach we have studied the effects of
 strong magnetic fields on the equation of state of  beta-equilibrium and
 charge neutrality matter relevant to neutron stars. The anomalous
 magnetic moments of both nucleons and electrons are incorporated in the
 model covariantly. We present the analytical expressions of the Dirac
 spinors under a uniform magnetic field. Numerical results show that if the
 magnetitude of the magnetic field is on the order of the surface field
 of neutron stars,
 the effects of the AMM are negligible. The EOS deviates from the field free
 case only at very low density region. If a much larger field is considered,
 the AMM plays a significant role so that the EOS becomes quite stiffer
 compared to the nonmagnetic case. Consequently, if the interior magnetic
 field of a neutron star is in the same level as the surface field
 inferred from pulsars, it may not cause evident influences on the star mass
 and radius as well as the maximum mass of star sequence. If ultra-strong
 fields do exist in neutron stars, considerable effects can be observed.
 We assume the magnetic field varies from the surface to the center
  of a neutron star. For the surface field we take the values inferred from
 pulsars. Strong fields up to
 $10^{18}$ $\sim$ $10^{19}$ G have been considered for the center of
 the neutron star. With the employed parametrization of the
 field changing with the density, the maximum mass could be enhanced
  by about 40\%.
 The central density of the star is in turn suppressed dramatically. For
 two equal-mass magnetic and nonmagnetic stars, the radius of magnetic one
 can be larger by about 10\% $\sim$ 20\% than the nonmagnetic one.

 It would be interesting to check whether the enhancement of the maximum
 mass induced by strong internal fields can be balanced by the hyperon degrees
 of freedom which was known to decrease the maximum mass. Since the effects of
 star rotations increase the star radii, it is necessary to study the
 effects of rotations and magnetic fields simultaneously.
 Theoretically one should solve the coupled Einstein-Maxwell equations
 for axisymmetric configuration  with the effects of
 magnetic fields taken into account both in the source term of the
 gravitational equation of general relativity and the nuclear equation
 of state. If in future
 astrophysical observations one can catch two pulsars with similar
 rotation periods and masses but quite different radii, a possible
 explanation is that the star having larger radius may contain a strong
 interior magnetic field.

 Now let us discuss several issues involved for the description of an
 electron in intense magnetic fields. It has long been recognized that
 the ground-state energy of an electron may be shifted after taking into
 account the AMM term. In previous calculations (O'Connell 1968;
 Chiu \& Canuto 1968) one
 thought that the Landau levels always start from $\nu = 0$. Thus, the
 ground-state energy of an electron in the presence of the magnetic field
 can be written as
 \begin{equation}
 E_{0} = m_{e} - \frac{\alpha}{2\pi}\mu_{B} B.
 \end{equation}
 The vacuum becomes unstable with respect to the electron-positron pair
 creations when $B > 7.6 \times 10^{16}$ G, which may have dramatic
 astrophysical consequences. As pointed out in the
 appendix, the Landau levels start from $\nu = 1$ for spin-up particles
 since the wave functions vanish at $\nu = 0$. It causes the ground-state
 energy to the form of
  \begin{equation}
 E_{0} = \sqrt{m_{e}^{2} + 2eB } - \frac{\alpha}{2\pi}\mu_{B} B.
 \label{cripair}
  \end{equation}
The critical field for pair creations now turns out to be
  $2.6 \times 10^{20}$ G. However, Eq. (\ref{cripair}) is valid at the level of
 the anomalous magnetic moment term (Jancovici 1969). Higher-order terms,
e.g. radiative correction due to vacuum polarization effect (Schwinger 1951; 1973),
 may become effective at such large field. The incorporation of these terms
 in the present model is a problem to be addressed in future studies.

 \end{sloppypar}
\setcounter{equation}{0}
\renewcommand{\theequation}{A\arabic{equation}}
  \begin{center}
 {\bf APPENDIX A}
  \end{center}
  \begin{sloppypar}
 In this appendix we derive the Dirac spinors and energy spectra
of nucleons in the presence of a constant uniform magnetic field along the $z$ axis.
 The Dirac equation for a free nucleon which has an anomalous magnetic moment
 $\mu_{N}$ in an external magnetic field can be written as
  \begin{equation}
 \left[ i\gamma_{\mu}\partial^{\mu}
   - e \frac{1+\tau_{0}}{2} \gamma_{\mu}A^{\mu} -\frac{1}{4}\kappa_{b}\mu_{N}
   \sigma_{\mu\nu}F^{\mu\nu}
  - M_{N} \right] \psi =0.  \label{diracfree}
  \end{equation}
 If one drops the term concerning the AMM, the solutions for neutrons are just the
 conventional Dirac spinors (Greiner 1990; Weinberg 1995).
 The Dirac theory for free electrons
 in a homogeneous magnetic field was first investigated
  by Rabi (1928). The
wave functions for charged particles without the inclusion of the AMM have been
 studied by several authors (Kobayashi \& Sakamoto 1983; Das \& Hott 1996).
 Johnson and Lippmann (1950) considered
  the inclusion of the AMM in the
 Dirac equation within a noncovariant description. A covariant energy spectrum
 was discussed by Vshivtsev and Serebryakova (1994).
 Recently, Broderick, Prakash and Lattimer (Paper II)
  derived the spinors and energy spectra for baryons in the
Dirac representation. However, their formulae are quite complicated which leads
to a lack of analytical expressions for the wave functions. Here we resolve the
problem in the chiral representation. The employed $\gamma$-matrices then become
 (Itzykson \& Zuber 1980)
   \begin{eqnarray}
 \gamma^{0}= \left( \begin{array}{cc}  0  &  -I \\  -I & 0 \end{array} \right),
 \qquad \qquad
 \mbox{\boldmath $\alpha$} = \left( \begin{array}{cc}
 \mbox{\boldmath $\sigma$}  &  0 \\  0 & - \mbox{\boldmath $\sigma$}
 \end{array} \right),  \\
 \mbox{\boldmath $\gamma$} = \left( \begin{array}{cc}
 0  & \mbox{\boldmath $\sigma$}  \\  - \mbox{\boldmath $\sigma$} & 0
 \end{array} \right),
 \qquad \qquad
 \gamma_{5}= \left( \begin{array}{cc}  I  &  0 \\  0 & -I \end{array} \right),
   \end{eqnarray}
 where $\mbox{\boldmath $\sigma$}$ is the Pauli matrix.
 Equation~(\ref{diracfree}) can be rewritten as
   \begin{equation}
 i \frac{\partial}{\partial t}\psi = \left[ -i \mbox{\boldmath $\alpha$} \cdot
   \mbox{\boldmath $\nabla$} - e\frac{1+\tau_{0}}{2}\alpha_{2} B x +\beta M_{N}
   + \frac{1}{4}\beta \kappa_{b}\mu_{N} \sigma_{\mu\nu}F^{\mu\nu} \right]
   \psi .  \label{diraceig}
   \end{equation}
 In the following we define $\Delta = - \frac{1}{2}\kappa_{b} \mu_{N}B$ and
 consider the cases of protons and neutrons individually.

   \begin{center}
   Protons
   \end{center}
   Let us specify the wave functions of protons as
 \begin{equation}
   \psi(X)=e^{-iEt + ip_{y}y + ip_{z}z}\phi_{p}(p_{y},p_{z},x).
 \end{equation}
 In the static system we obtain the eigenequation
 \begin{eqnarray}
 \left( \begin{array}{cccc}
   p_{z}-E       &    \xi_{+}    &  -M_{N} + \Delta   &  0  \\
   \xi_{-}       &    -p_{z}-E   &       0            & -M_{N} - \Delta \\
 -M_{N}+\Delta   &       0       &  -p_{z} - E        &  -\xi_{+}        \\
      0          & -M_{N}-\Delta &      -\xi_{-}      &   p_{z} - E \end{array}
      \right) \left( \begin{array}{c}
   \phi_{p}^{(1)} \\  \phi_{p}^{(2)} \\ \phi_{p}^{(3)} \\ \phi_{p}^{(4)}
   \end{array} \right) =0 . \label{eigfpro}
 \end{eqnarray}
 Here we have defined $\xi_{\pm} \equiv -i\partial_{x} \mp i(p_{y} -eBx)$.
 Provided $eB > 0$, one can introduce the eigenfunction of $\xi_{+}$, $\xi_{-}$
 \begin{eqnarray}
 I_{\nu,p_{y}}(x) &=& \left( \frac{eB}{\pi} \right) ^{1/4} \exp \left[
    -\frac{1}{2}eB \left( x-\frac{p_{y}}{eB} \right)^{2} \right]
    \frac{1}{\sqrt{\nu !}}H_{\nu} \left[ \sqrt{2eB} \left( x - \frac{p_{y}}
    {eB} \right) \right], \nonumber \\
    && (\nu = 0, 1, 2, ...)
 \end{eqnarray}
 where $H_{\nu}(x)$ is the Hermite polynomial defined by
  \begin{equation}
 H_{\nu}(x)= (-1)^{\nu} \exp \left(\frac{x^{2}}{2}\right) \frac{d^{\nu}}{dx^{\nu}}
   \exp \left( -\frac{x^{2}}{2} \right) .
  \end{equation}
 $I_{\nu,p_{y}}(x)$ is normalized as
  \begin{eqnarray}
 && \int\, dx I_{\nu,p_{y}}(x) I_{\mu,p_{y}}(x) = \delta_{\nu\mu}, \\
 && \sum_{\nu=0}^{\infty} I_{\nu,p_{y}}(x) I_{\nu,p_{y}}(x^{\prime})
    = \delta(x-x^{\prime}) .
  \end{eqnarray}
 It satisfies the following relations:
  \begin{eqnarray}
 && \xi_{-} I_{\nu,p_{y}}(x) = -i \sqrt{2eB\nu} I_{\nu -1, p_{y}}(x),
    \qquad (I_{-1,p_{y}}(x)=0) \\
 && \xi_{+} I_{\nu,p_{y}}(x) = i \sqrt{2eB(\nu + 1)} I_{\nu +1, p_{y}}(x).
  \end{eqnarray}
 The eigenvalues and eigenfunctions of Eq.~(\ref{eigfpro}) can be deduced in a
 standard way by performing matrix calculations. The energy spectra of protons
 are
   \begin{equation}
 \left( E_{\nu,S}^{p} \right)_{+}= E_{S}^{\nu}, \qquad \qquad
 \left( E_{\nu,S}^{p} \right)_{-}= - E_{S}^{\nu},
   \end{equation}
 with
  \begin{equation}
 E_{S}^{\nu} = \left[ \left( \sqrt{M_{N}^{2} + 2eB\nu } + S\Delta \right) ^{2}
   + p_{z}^{2} \right] ^{1/2},
  \end{equation}
 here $S=\pm 1$ for the spin-up and spin-down particles. The respective
 eigenfunctions are as follows:
  \begin{eqnarray}
 \psi_{1}(X) &=& e^{-i E_{+1}^{\nu}t + ip_{y}y + ip_{z}z} \sqrt{ \frac{
    E_{+1}^{\nu} + p_{z}}{2E_{+1}^{\nu}} } \left[ 1 + \frac{2eB\nu}
    { \left( \sqrt{M_{N}^{2} + 2eB\nu} + M_{N} \right) ^{2} } \right] ^{-1/2}
    \nonumber \\
    && \times \left( \begin{array}{c}
   - \frac{i\sqrt{2eB\nu}}{\sqrt{M_{N}^{2} + 2eB\nu} + M_{N} }I_{\nu,p_{y}}(x) \\ \\
    - \frac{\sqrt{M_{N}^{2} + 2eB\nu} + \Delta}{E_{+1}^{\nu} + p_{z}}
    I_{\nu -1, p_{y}}(x) \\  \\
    - \frac{i\sqrt{2eB\nu} \left( \sqrt{M_{N}^{2} + 2eB\nu} + \Delta \right) }
    { \left( E_{+1}^{\nu} + p_{z} \right) \left( \sqrt{M_{N}^{2} + 2eB\nu}
    + M_{N} \right) } I_{\nu,p_{y}}(x) \\ \\
    I_{\nu -1, p_{y}}(x)  \end{array} \right) ,   \\
 \psi_{2}(X) &=& e^{-i E_{-1}^{\nu}t + ip_{y}y + ip_{z}z} \sqrt{ \frac{
    E_{-1}^{\nu} + p_{z}}{2E_{-1}^{\nu}} } \left[ 1 + \frac{2eB\nu}
    { \left( \sqrt{M_{N}^{2} + 2eB\nu} - M_{N} \right) ^{2} } \right] ^{-1/2}
    \nonumber \\
    && \times \left( \begin{array}{c}
    \frac{i\sqrt{2eB\nu}}{\sqrt{M_{N}^{2} + 2eB\nu} - M_{N} }I_{\nu,p_{y}}(x) \\ \\
    \frac{\sqrt{M_{N}^{2} + 2eB\nu} - \Delta}{E_{-1}^{\nu} + p_{z}}
    I_{\nu -1, p_{y}}(x) \\ \\
    - \frac{i\sqrt{2eB\nu} \left( \sqrt{M_{N}^{2} + 2eB\nu} - \Delta \right) }
    { \left( E_{-1}^{\nu} + p_{z} \right) \left( \sqrt{M_{N}^{2} + 2eB\nu}
    - M_{N} \right) } I_{\nu,p_{y}}(x) \\ \\
    I_{\nu -1, p_{y}}(x) \end{array} \right) ,  \\
 \psi_{3}(X) &=& e^{-i (- E_{+1}^{\nu}) t + ip_{y}y + ip_{z}z} \sqrt{ \frac{
    E_{+1}^{\nu} + p_{z}}{2E_{+1}^{\nu}} } \left[ 1 + \frac{2eB\nu}
    { \left( \sqrt{M_{N}^{2} + 2eB\nu} + M_{N} \right) ^{2} } \right] ^{-1/2}
    \nonumber \\
    && \times \left( \begin{array}{c}
    - \frac{i\sqrt{2eB\nu} \left( \sqrt{M_{N}^{2} + 2eB\nu} + \Delta \right) }
    {\left( E_{+1}^{\nu} + p_{z} \right) \left( \sqrt{M_{N}^{2} + 2eB\nu}
    + M_{N} \right) } I_{\nu,p_{y}}(x) \\ \\
    I_{\nu -1, p_{y}}(x) \\ \\
    \frac{i\sqrt{2eB\nu} }{\sqrt{M_{N}^{2} + 2eB\nu} + M_{N} }I_{\nu,p_{y}}(x) \\ \\
    \frac{\sqrt{M_{N}^{2} + 2eB\nu} + \Delta }{E_{+1}^{\nu} + p_{z}}
    I_{\nu -1, p_{y}}(x) \end{array} \right) , \\
 \psi_{4}(X) &=& e^{-i (-E_{-1}^{\nu})t + ip_{y}y + ip_{z}z} \sqrt{ \frac{
    E_{-1}^{\nu} + p_{z}}{2E_{-1}^{\nu}} } \left[ 1 + \frac{2eB\nu}
    { \left( \sqrt{M_{N}^{2} + 2eB\nu} - M_{N} \right) ^{2} } \right] ^{-1/2}
    \nonumber \\
    && \times \left( \begin{array}{c}
    \frac{i\sqrt{2eB\nu} \left( \sqrt{M_{N}^{2} + 2eB\nu} - \Delta \right) }
    {\left( E_{-1}^{\nu} + p_{z} \right) \left( \sqrt{M_{N}^{2} + 2eB\nu}
    - M_{N} \right) } I_{\nu,p_{y}}(x) \\ \\
    - I_{\nu -1,p_{y}}(x) \\ \\
    \frac{i\sqrt{2eB\nu}}{\sqrt{M_{N}^{2} + 2eB\nu} - M_{N} }I_{\nu,p_{y}}(x) \\ \\
    \frac{\sqrt{M_{N}^{2} + 2eB\nu} - \Delta}{E_{-1}^{\nu} + p_{z}}
    I_{\nu -1,p_{y}}(x) \end{array} \right) .
  \end{eqnarray}
 One can easily check that $\psi_{i}(i=1,4)$ forms a complete orthogonal set.
 Note that the Landau levels start at $\nu = 0$ for spin-down particles and
 $\nu =1$ for spin-up particles since $\psi_{1}$ and $\psi_{3}$ vanish at
 $\nu = 0$. As pointed out in Sect.~II, here the so-called spin up and spin down
 are just relative notes for convenience of description.

   \begin{center}
   Neutrons
   \end{center}
 The wave functions of neutrons can be specified as
   \begin{equation}
 \psi(X) = e^{-iEt + i{\bf p}\cdot {\bf x}} \phi_{n}({\bf p}).
   \end{equation}
 Inserting it into Eq.~(\ref{diraceig}) we have the following eigenequation:
  \begin{eqnarray}
 \left( \begin{array}{cccc}
   p_{z}-E       &    p_{x}-ip_{y}    &  -M_{N} + \Delta   &  0  \\
 p_{x} + ip_{y}  &    -p_{z}-E   &       0            & -M_{N} - \Delta \\
 -M_{N}+\Delta   &       0       &  -p_{z} - E        &  -(p_{x} - ip_{y})    \\
      0          & -M_{N}-\Delta &  -(p_{x} + ip_{y}) &   p_{z} - E \end{array}
      \right) \left( \begin{array}{c}
   \phi_{n}^{(1)} \\  \phi_{n}^{(2)} \\ \phi_{n}^{(3)} \\ \phi_{n}^{(4)}
   \end{array} \right) =0 . \label{eigfneu}
  \end{eqnarray}
 Through solving the above matrix equation we obtain the energy spectra of
 neutrons as
  \begin{equation}
 \left( E_{S}^{n} \right) _{+} = E_{S}, \qquad \qquad
 \left( E_{S}^{n} \right)_{-} = -E_{S},
  \end{equation}
 with
  \begin{equation}
 E_{S}= \left[ \left( \sqrt{p_{x}^{2} + p_{y}^{2} + M_{N}^{2} } + S\Delta
   \right) ^{2} + p_{z}^{2} \right]^{1/2} .
  \end{equation}
 The corresponding eigenfunctions read as
  \begin{eqnarray}
 \psi_{1}(X) &=& e^{-i E_{+1}t + i{\bf p}\cdot {\bf x}} \sqrt{ \frac{
    E_{+1} + p_{z}}{2E_{+1}}} \left[ 1 + \frac{p_{x}^{2} + p_{y}^{2}}
    { \left( \sqrt{M_{N}^{2} + p_{x}^{2} + p_{y}^{2}}
    + M_{N} \right) ^{2} } \right] ^{-1/2}
    \nonumber \\
    && \times \left( \begin{array}{c}
    - \frac{p_{x} - ip_{y}}{\sqrt{M_{N}^{2} + p_{x}^{2} + p_{y}^{2} }+M_{N}} \\ \\
    - \frac{\sqrt{M_{N}^{2} + p_{x}^{2} + p_{y}^{2}} + \Delta}
    {E_{+1} + p_{z}} \\ \\
    - \frac{\left( p_{x} - ip_{y} \right) \left( \sqrt{M_{N}^{2} + p_{x}^{2}
    + p_{y}^{2} } + \Delta \right) }{\left( E_{+1} + p_{z} \right)
    \left( \sqrt{M_{N}^{2} + p_{x}^{2} + p_{y}^{2}} + M_{N} \right) } \\
    1
    \end{array} \right) , \\
 \psi_{2}(X) &=& e^{-i E_{-1}t + i{\bf p}\cdot {\bf x}} \sqrt{ \frac{
    E_{-1} + p_{z}}{2E_{-1}}} \left[ 1 + \frac{p_{x}^{2} + p_{y}^{2}}
    { \left( \sqrt{M_{N}^{2} + p_{x}^{2} + p_{y}^{2}}
    - M_{N} \right) ^{2} } \right] ^{-1/2}
    \nonumber \\
    && \times \left( \begin{array}{c}
    \frac{p_{x} - ip_{y}}{\sqrt{M_{N}^{2} + p_{x}^{2} + p_{y}^{2}} - M_{N}} \\ \\
    \frac{\sqrt{M_{N}^{2} + p_{x}^{2} + p_{y}^{2}} - \Delta }{E_{-1} + p_{z}} \\ \\
    - \frac{\left( p_{x} - ip_{y} \right) \left( \sqrt{M_{N}^{2} + p_{x}^{2}
    + p_{y}^{2}} - \Delta \right) }{ \left( E_{-1} + p_{z} \right)
    \left( \sqrt{M_{N}^{2} + p_{x}^{2} + p_{y}^{2}} - M_{N} \right) } \\
    1  \end{array} \right) , \\
 \psi_{3}(X) &=& e^{-i (-E_{+1})t + i{\bf p}\cdot {\bf x}} \sqrt{ \frac{
    E_{+1} + p_{z}}{2E_{+1}}} \left[ 1 + \frac{p_{x}^{2} + p_{y}^{2}}
    { \left( \sqrt{M_{N}^{2} + p_{x}^{2} + p_{y}^{2}}
    + M_{N} \right) ^{2} } \right] ^{-1/2}
    \nonumber \\
    && \times \left( \begin{array}{c}
    - \frac{ \left( p_{x} - ip_{y} \right) \left( \sqrt{M_{N}^{2} + p_{x}^{2}
    + p_{y}^{2}} + \Delta \right) }{ \left( E_{+1} + p_{z} \right)
    \left( \sqrt{M_{N}^{2} + p_{x}^{2} + p_{y}^{2}} + M_{N} \right) } \\
    1  \\
    \frac{p_{x} - ip_{y}}{\sqrt{M_{N}^{2} + p_{x}^{2} + p_{y}^{2}} + M_{N} } \\ \\
    \frac{\sqrt{M_{N}^{2} + p_{x}^{2} + p_{y}^{2} } + \Delta }
    {E_{+1} + p_{z} } \end{array} \right) , \\
 \psi_{4}(X) &=& e^{-i (-E_{-1})t + i{\bf p}\cdot {\bf x}} \sqrt{ \frac{
    E_{-1} + p_{z}}{2E_{-1}}} \left[ 1 + \frac{p_{x}^{2} + p_{y}^{2}}
    { \left( \sqrt{M_{N}^{2} + p_{x}^{2} + p_{y}^{2}}
    - M_{N} \right) ^{2} } \right] ^{-1/2}
    \nonumber \\
    && \times \left( \begin{array}{c}
    \frac{\left( p_{x} - ip_{y} \right) \left( \sqrt{M_{N}^{2} + p_{x}^{2}
    + p_{y}^{2} } - \Delta \right) }{ \left( E_{-1} + p_{z} \right) \left( \sqrt{
    M_{N}^{2} + p_{x}^{2} + p_{y}^{2} } - M_{N} \right) } \\
    -1 \\
    \frac{p_{x} - ip_{y}}{ \sqrt{M_{N}^{2} + p_{x}^{2} + p_{y}^{2} } - M_{N} } \\ \\
    \frac{\sqrt{M_{N}^{2} + p_{x}^{2} + p_{y}^{2} }  - \Delta}
    {E_{-1} + p_{z} } \end{array} \right) .
  \end{eqnarray}
  Again, $\psi_{i}(i=1,4)$ is orthonormalized.

  \end{sloppypar}
 \vspace{0.5cm}
 \begin{sloppypar}
 \noindent {\bf Acknowledgments:} G.~Mao was financially supported by the
 STA foundation and in part supported by the National Natural Science
 Foundation of China under the grant 10275072. He thanks the Japan Atomic
 Energy Research Institute and the China Institute of Atomic Energy for
 local hospitality during visit.

 \end{sloppypar}

% \newpage

 \newpage
 \begin{figure}[htbp]
 \vspace{-10cm}
 \mbox{\hskip 4.8cm \psfig{file=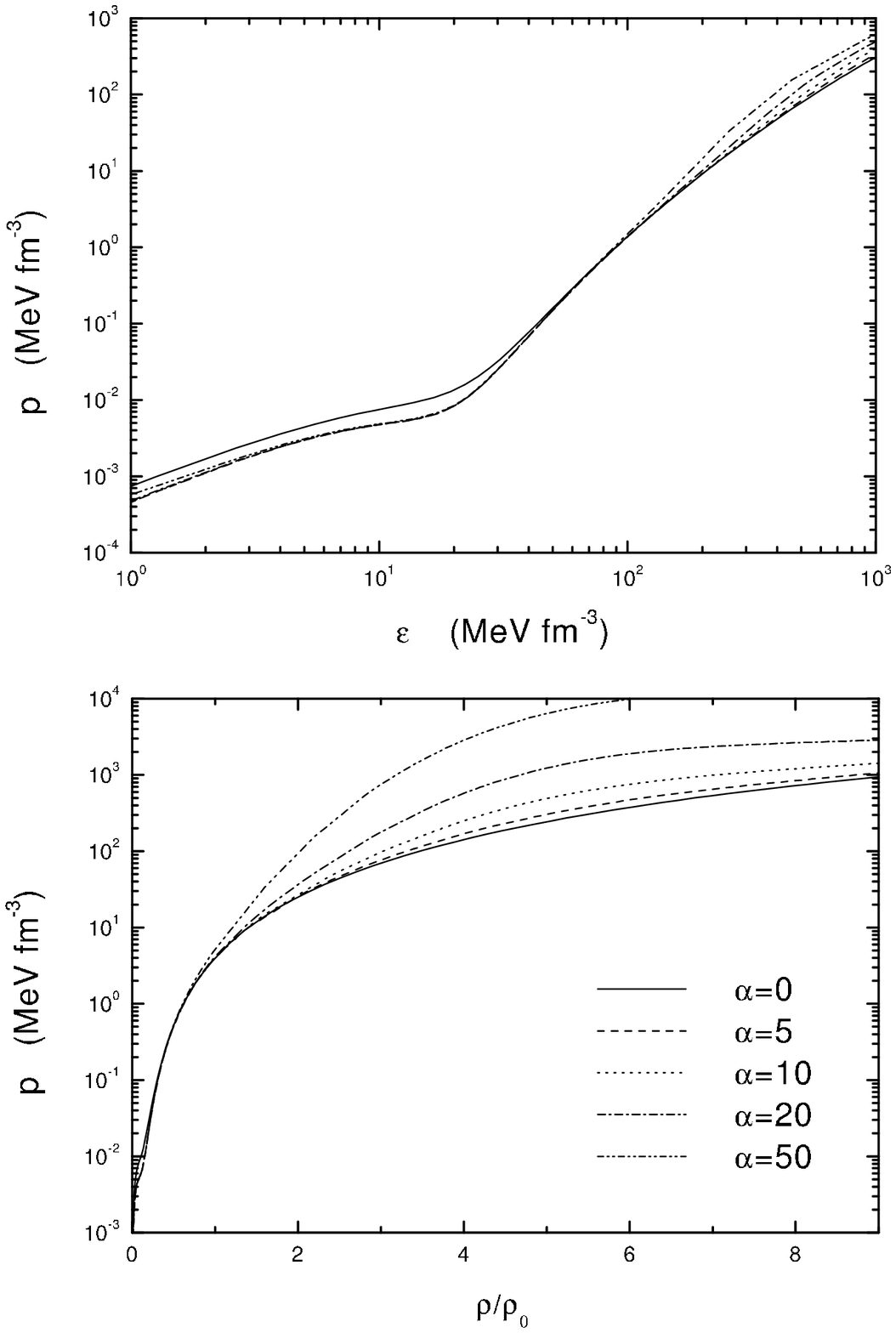,width=18cm,height=24cm,angle=0}}
 \vspace{2cm}
 \caption{The equation of state of neutron-star matter with strong magnetic
 fields. The upper panel gives the relationship between the pressure and
 total energy density while the lower panel displays the pressure as a
 function of density. Different curves correspond to different magnetic
 field strength as indicated in the figure and explained in the text.}
 \end{figure}

\newpage
 \begin{figure}[htbp]
 \vspace{3.0cm}
 \mbox{\hskip 3.0cm \psfig{file=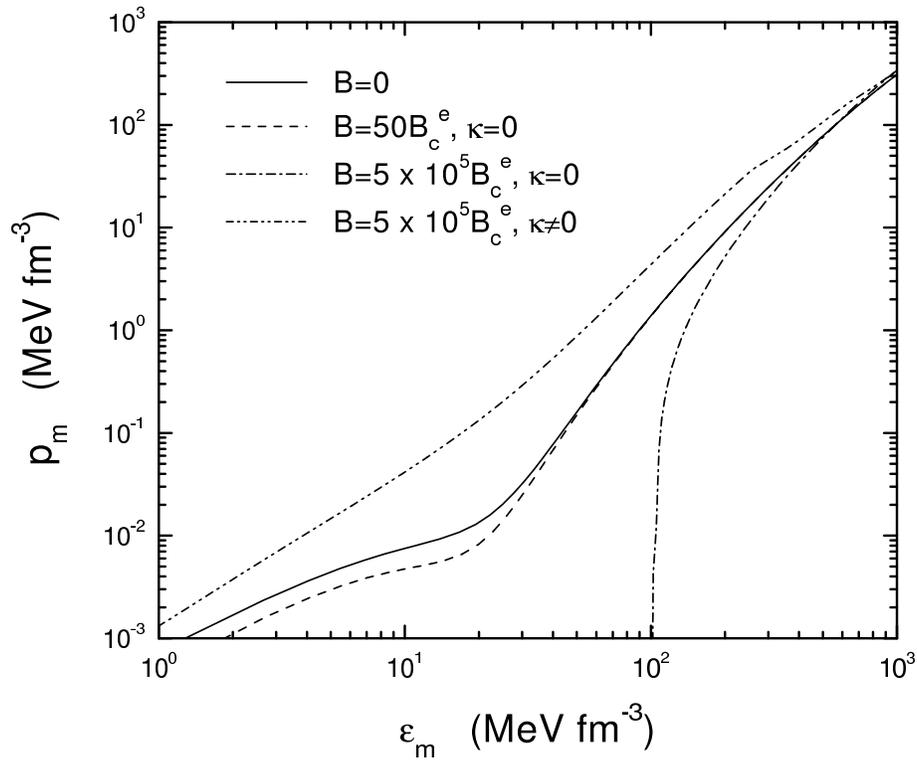,width=24cm,height=20cm,angle=-90}}
 \vspace{-11cm}
 \caption{The equation of state of neutron-star matter with a uniform
 magnetic field. The contributions of the magnetic field to the pressure
 and energy density are neglected. Different curves are related to
 different cases of magnetic fields $B$ and with or without the inclusion
 of the anomalous magnetic moments $\kappa$ as indicated in the figure.}
 \end{figure}

 \newpage
 \begin{figure}[htbp]
 \vspace{3.0cm}
 \mbox{\hskip 6.8cm \psfig{file=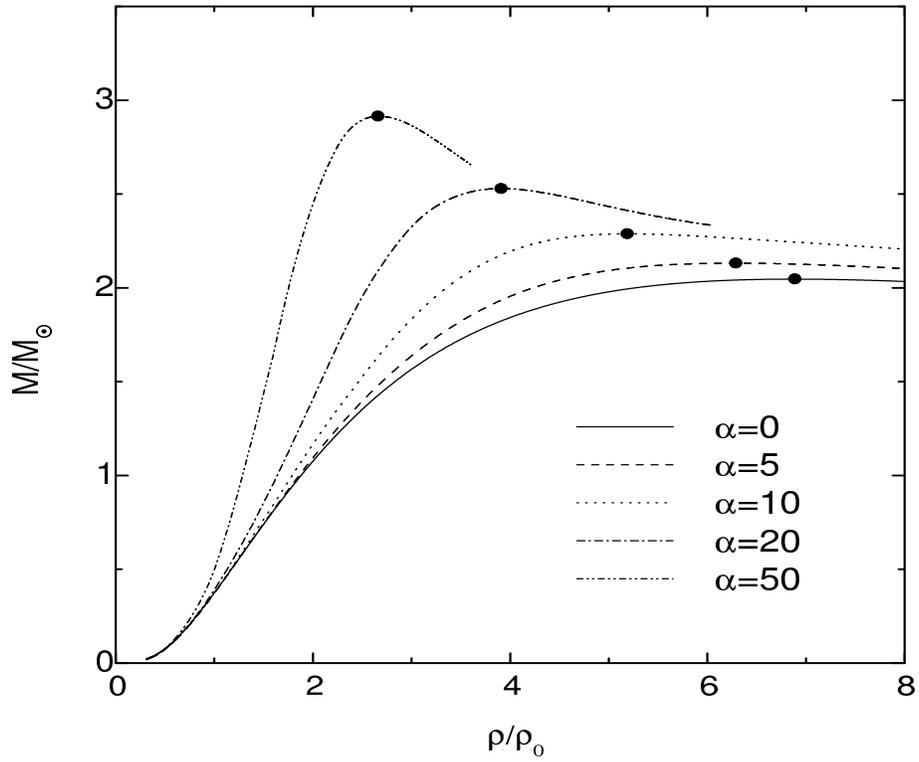,width=24cm,height=20cm,angle=-90}}
 \vspace{-11cm}
 \caption{The gravitational masses of neutron stars as a function of
 central densities, i.e., the sequences of stars under different interior
 magnetic fields as depicted by different curves in the figure.}
 \end{figure}

\newpage
 \begin{figure}[htbp]
 \vspace{3.0cm}
 \mbox{\hskip 4.8cm \psfig{file=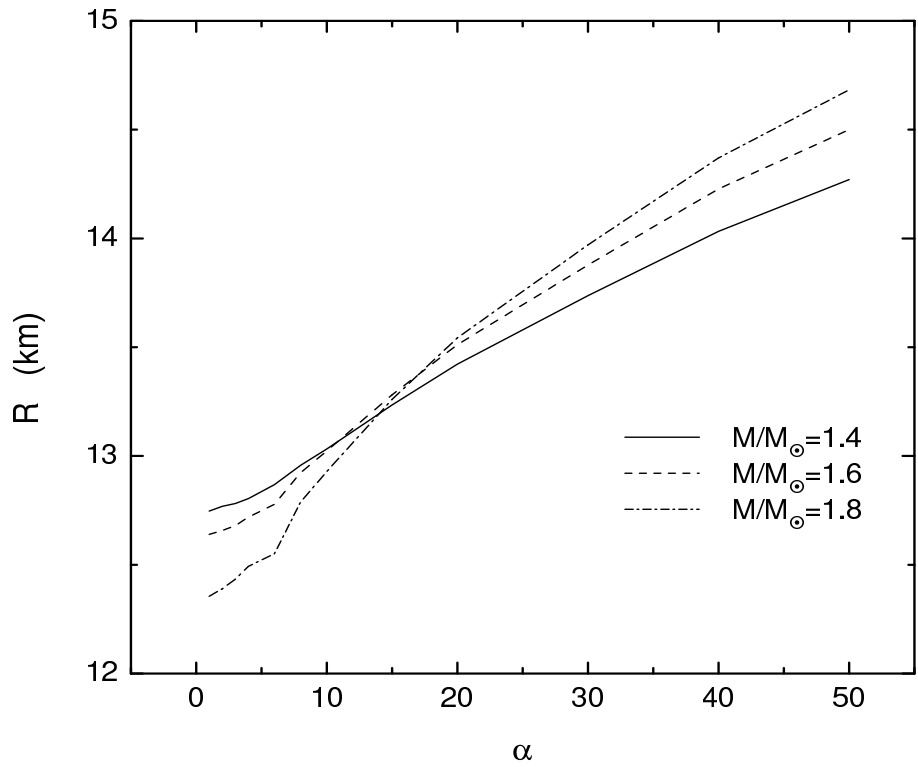,width=24cm,height=20cm,angle=-90}}
 \vspace{-11cm}
 \caption{The radii of neutron stars with equal gravitational masses as
 a function of interior magnetic field strength. Different curves correspond
 to different groups of star masses.}
 \end{figure}

\newpage
 \begin{figure}[htbp]
 \vspace{0.0cm}
 \mbox{\hskip -2.8cm \psfig{file=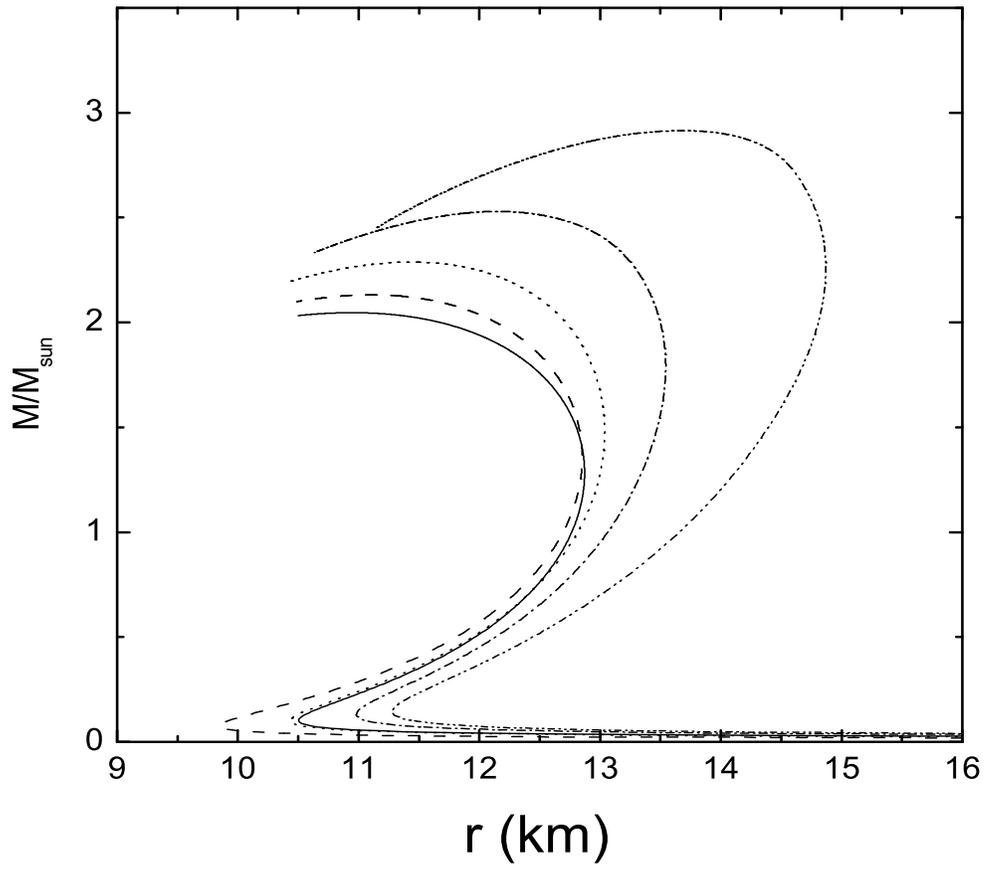,width=20cm,height=18cm,angle=0}}
 \vspace{-3cm}
 \caption{The mass-radius relations of neutron-star sequences. The different curves
 are related to the different interior magnetic fields as depicted in the legend
 of Fig.~3.}
 \end{figure}

\end{document}